\documentclass[final]{svjour2}
\usepackage{graphicx}
\usepackage{rotating}
\usepackage{amssymb}
\usepackage{mathptmx}
\usepackage[numbers]{natbib}
\makeatletter
\journalname{Journal of Low Temperature Physics}

\bibpunct{}{}{,}{s}{}{,}

\begin{document}

\newcommand{\hdblarrow}{H\makebox[0.9ex][l]{$\downdownarrows$}-}
\title{Andreev Reflections in Micrometer-Scale Normal-Insulator-Superconductor Tunnel Junctions\footnote[1]{Contribution of the US government; not subject to copyright in the United States.}}

\author{Peter J. Lowell \and Galen C. O'Neil \and Jason M. Underwood \and Joel N. Ullom}

\institute{National Institute of Standards and Technology\\ 325 Broadway MS 817.03 \\Boulder, CO 80305, USA\\
\email{peter.lowell@nist.gov}}

\date{21 October 2011}

\maketitle


\begin{abstract}

Understanding the subgap behavior of Normal-Insulator-Superconductor (NIS) tunnel junctions is important in order to be able to accurately model the thermal properties of the junctions.  Hekking and Nazarov \cite{hekking1994subgap} developed a theory in which NIS subgap current in thin-film structures can be modeled by multiple Andreev reflections.    In their theory, the current due to Andreev reflections depends on the junction area and the junction resistance area product.  We have measured the current due to Andreev reflections in NIS tunnel junctions for various junction sizes and junction resistance area products and found that the multiple reflection theory is in agreement with our data.

\end{abstract}


\section{Introduction}

Accurate modeling of the current-voltage (IV) characteristics of NIS junctions is required in order to use the junctions in applications such as primary thermometers or solid-state refrigerators.  The Bardeen-Cooper-Schrieffer (BCS) theory of superconductivity accurately predicts the IV characteristics of NIS junctions above the superconducting energy gap $\Delta$, while predicting almost no current when the junction is biased below the gap \cite{BCS,Tinkham}.  When NIS junctions are measured, subgap currents greater than BCS predictions are often measured.  This excess current can be explained by Andreev reflections \cite{Andreev,SaintJames}, where an electron (hole) in the normal metal is reflected from the NS interface as a hole (electon), which allows a Cooper pair to enter (leave) the superconductor.  In interfaces where the electrons and quasiparticles are in the ballistic regime and can be represented as a plane wave, the Andreev reflection is described by the Blonder-Tinkham-Klapwijk (BTK) theory\cite{BTKtheory}.  This excess power load in NIS junctions was modeled by Bardas and Averin \cite{BardasAverin}.  In realistic interfaces, the electrons and quasiparticles are no longer in the ballistic regime but behave diffusively, because they can reflect off the barrier and surrounding surfaces many times before tunneling, which will cause a higher current than predicted by the BTK theory.  Hekking and Nazarov developed a model to account for the extra current due to multiple Andreev reflections.  Rajauria et al. \cite{Rajauria2008} have made measurements of SINIS IV curves, and their subgap data agreed with Hekking and Nazarov's theory when they multiplied their data by a scaling factor of 1.37 \cite{Rajauria2}.  However, Rajauria et al. preformed measurements on only a single junction size and a single oxidation thickness.  In order to understand  NIS junctions, it must be understood how junction area and interface resistance affect the Andreev current.  In this paper, we provide a more robust test of Hekking and Nazarov's theory by comparing it with measurements of multiple NIS devices with different junction areas and resistance area products.

\section{Theory}

BCS theory predicts that current will flow through an NIS junction at voltage bias greater than the superconducting energy gap.  When the bias voltage is less than the superconducting gap, little current should flow because electron tunneling is limited by the density of states in the superconductor.  However, a current below the gap is possible from mechanisms such as Andreev reflections.  Hekking and Nazarov predict an additional current below the subgap caused by Andreev reflections:

 \begin{equation}
\ I_{Andreev} = \frac{\hbar}{e^{3}R^{2}_{n}S\nu_{N}d_{N}} \mathrm{tanh}(eV/2T)+ \frac{\hbar}{e^{3}R^{2}_{n}S\nu_{S}d_{S}} \frac{eV}{2 \pi \Delta \sqrt{1-eV/\Delta}}
\label{eq:andreev}
\end{equation}

\noindent where $e$ is the electron charge, $R_{n}$ is the normal state resistance of the junction, $S$ is the junction area, $\nu_{N,S}$ is the density of states of the normal metal (superconductor), $d_{N,S}$ is the thickness of the normal metal (superconductor), $V$ is the voltage bias and $T$ is the temperature.  By rearranging Eq. \ref{eq:andreev}, we find that the dimensionless quantity $eIR_{n}/\Delta$ scales inversely with the resistance area product.

Equation \ref{eq:andreev} is valid as long as the junction dimensions are larger than the coherence length in each metal.  The coherence length of the superconductor is given by $\xi_{S} = \sqrt{\hbar D_{S}/\Delta}$, and the coherence length of the normal metal is given by $\xi_{N} = \sqrt{\hbar D_{N}/k_{b} T}$, where $D_{S,N}$ is the diffusion constant for the superconductor and the normal metal, respectively.  We measured the resistivity of our superconductor, Al, to be $\rho_{Al}$ = 0.005 $\Omega \times \mu$m, and measured the resistivity of our normal metal, AlMn \cite{AlMn}, to be $\rho_{AlMn}$ = 0.10 $\Omega \times \mu$m. Using these values, we calculated the coherence lengths to be $\xi_{S}$ = 444 nm and $\xi_{N}$ = 571 nm.  Our smallest junction dimension is 2 $\mu$m, which is greater than both coherence lengths.  Therefore, it is valid to use Eq. \ref{eq:andreev} to model our junctions.



\section{Experimental Details}

\begin{figure}
\begin{center}
\includegraphics[%
  width=0.65\linewidth,
  keepaspectratio]{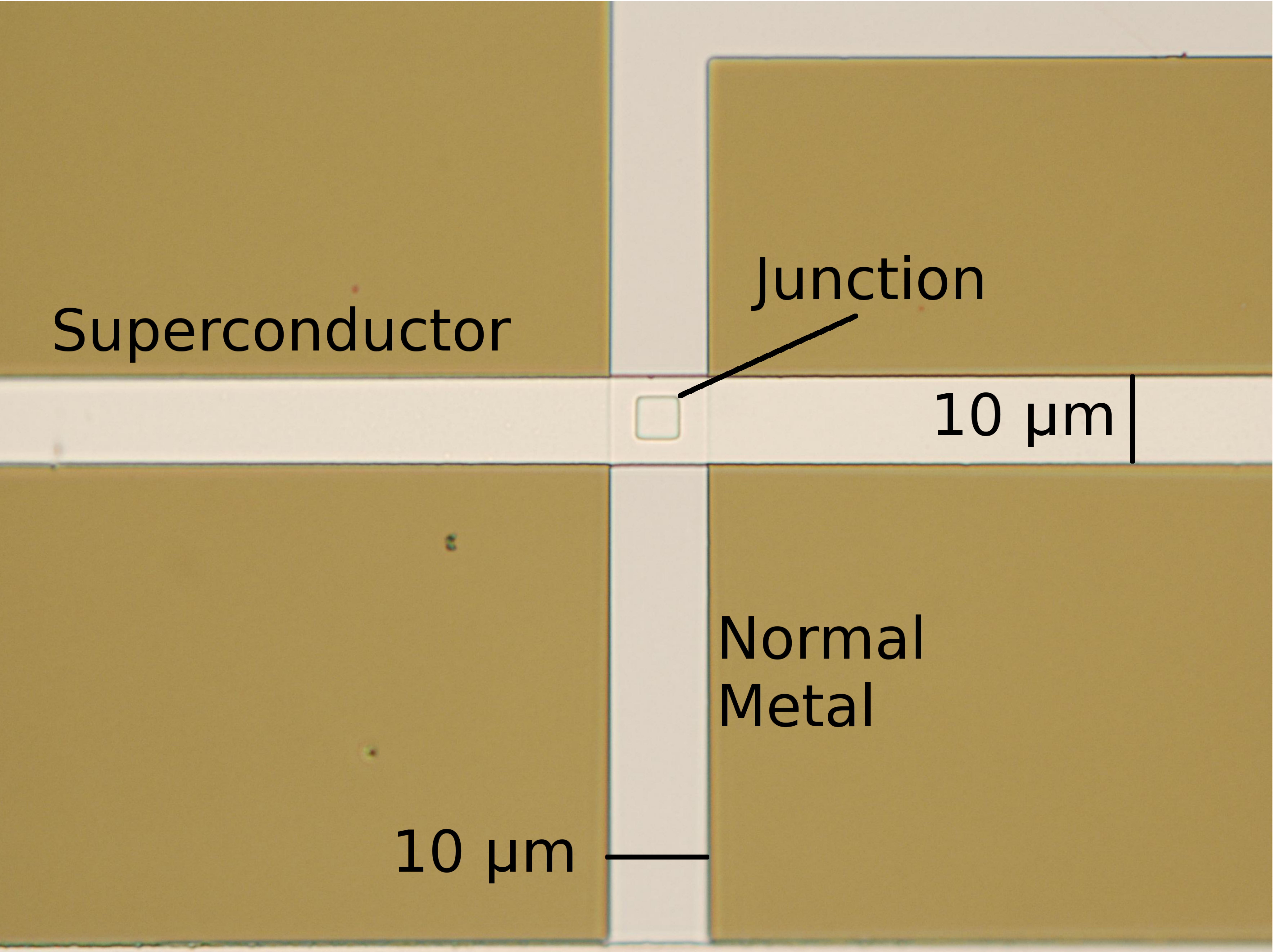}
\end{center}
\caption{(Color online) Optical image of a device used in our measurements.  The junction is made by overlapping normal metal and superconducting wires.  The junction area is defined by a via in a SiO$_{2}$ layer between the two metal layers.}
\label{fig:junction}
\end{figure}

Figure \ref{fig:junction} shows a typical device used in this experiment.   The junctions were created by intersecting normal metal and superconducting wires, and the junction area is defined by a via in a layer of SiO$_{2}$ separating the two metal layers.  The junctions were fabricated on a Si wafer by first sputter depositing 30 nm of Al doped with Mn to 4000 ppm by atomic percent.  The AlMn was patterned using standard photolithographic techniques and was etched in an acid bath.  A 90 nm thick layer of SiO$_{2}$ was deposited by use of plasma-enhanced chemical vapor deposition (PECVD) and vias were created by use of a plasma etch to define the junction area.  The devices were ion milled to remove any native oxide from the normal metal, and then exposed to oxygen to form the insulating layer.  A resistance area product, $R_{SP} \equiv R_{n} \times S$, of 30 $\Omega \mu$m$^{2}$ was created by exposing the devices to 0.1 torr-s of oxygen and a $R_{SP}$ of 200  $\Omega \mu$m$^{2}$ was created by exposing the devices to 42 torr-s of oxygen.  Finally, the superconducting Al counter electrode was sputter deposited and then wet etched by use of standard photolithographic techniques.  The 30 $\Omega \mu$m$^{2}$ devices were fabricated on top of a 150 nm layer of SiO$_{2}$ deposited by PECVD to increase the quality of the junctions, but no observable difference was measured between these junctions and junctions that were fabricated on just the thermal oxide.

The junctions were then screened to determine their quality before measurements were performed.  To determine the quality of our junctions, we use the quality factor $Q$, where $Q \equiv R_{leak}/R_{{n}}$.  The leakage resistance, $R_{leak}$, is defined as the highest resistance of the junction in the subgap.  Both wafers produced devices with a $Q \approx 2000$.  


For the experiment, we measured four devices, two from each wafer we fabricated.  We measured devices 3 $\mu$m by 3 $\mu$m and  4 $\mu$m by 4 $\mu$m from the wafer with a $R_{SP}$ = 200  $\Omega \mu$m$^{2}$,  and devices 2 $\mu$m by 2 $\mu$m and 3 $\mu$m by 3 $\mu$m  from the wafer with a $R_{SP}$ = 30 $\Omega \mu$m$^{2}$.  Devices from the same wafer were chosen from the same chip to make sure that the junction properties, such as the metal thicknesses and $R_{SP}$, were as similar as possible.  

\begin{table}
  \begin{tabular}{ | c | c | c | c | c | c | c|}
    \hline
    Device & Area ($\mu$m$^{2}$) & R$_{n} $ ($\Omega$) & $R_{SP}$ ($\Omega \mu$m$^{2}$)&d$_{N}$ (nm) & d$_{S}$ (nm)& $\Delta$ ($\mu$eV) \\ \hline
    1 & 16 & 11.2 & 179 &30  & 525 & 185 \\ \hline 
    2 & 9 & 21.5 & 193 &30  & 525 & 185 \\ \hline 
    3 & 9 & 2.98 & 27 &31  & 230 & 185 \\ \hline 
    4 & 4 & 8.25 & 33 &31 & 230 & 185  \\ \hline 
  \end{tabular}
  \label{tab:properties}
  \caption{Measured values of device parameters used in our experiment.}
\end{table}

IV measurements of the devices were made by four wire measurements in an adiabatic demagnetization refrigerator at 100 mK.  Current biasing was accomplished with a low-noise voltage source in series with a 10 M$\Omega$ resistor.  Table 1 shows the properties of the devices that were measured in this experiment.  The normal state resistance was measured from the differential resistance of the devices.  The normal metal and superconductor thicknesses were measured for each device by use of a profilometer.

\section{Results}

\begin{figure}
\begin{center}
\includegraphics[%
  width=.65\linewidth,
  keepaspectratio]{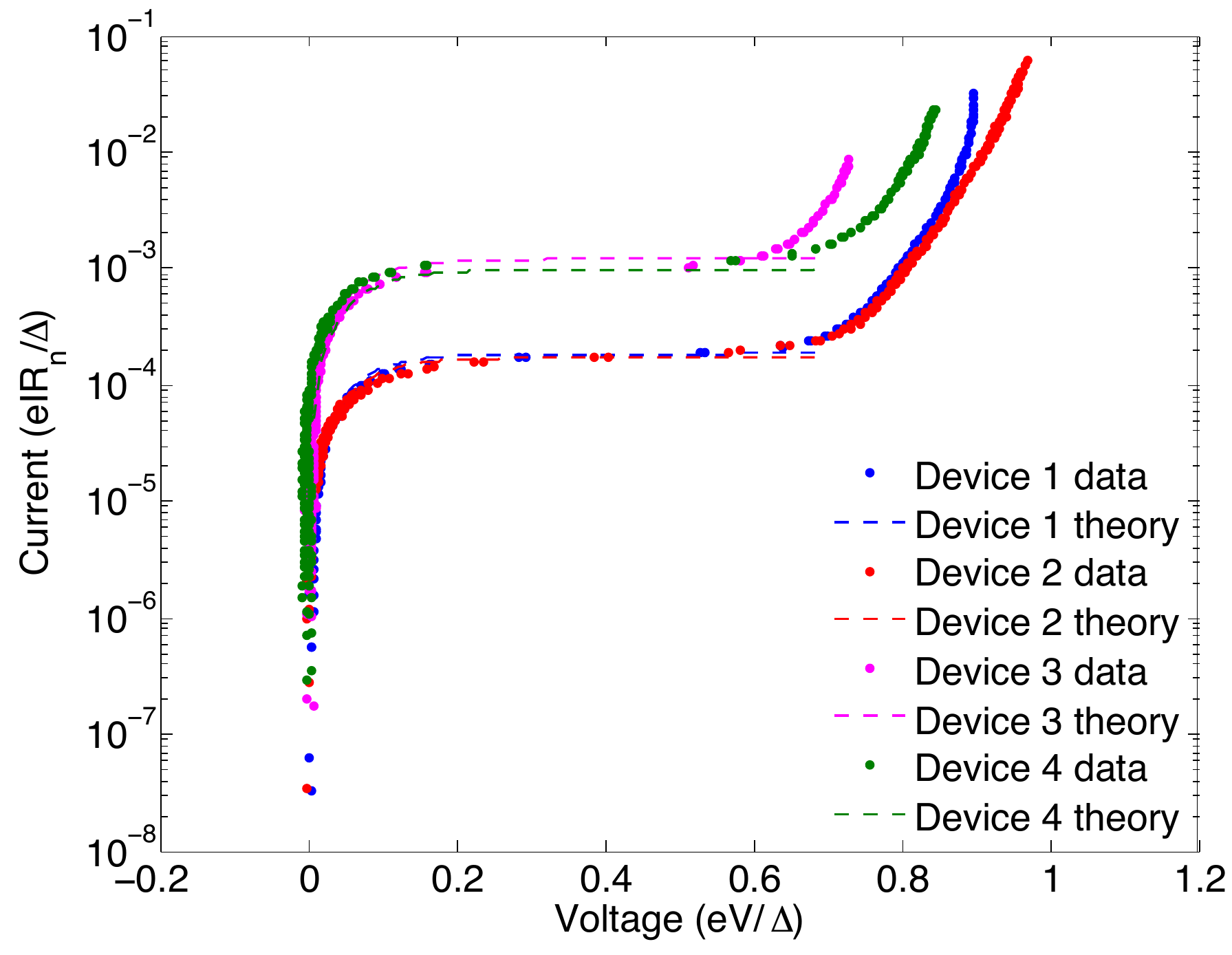}
\end{center}
\caption{(Color online) Current vs voltage for the four devices that we measured.  The data are represented by points and the theory as dashed lines.  The error bars for the uncertainty in the measurement are smaller than the data points.  The theory was calculated using the device parameters shown in Table 1.  The data are in excellent agreement with the theory at lower voltage biases.  At higher biases, a detailed thermal model is needed to match the theory with data.}
\label{fig:Andreev}
\end{figure}

\begin{figure}
\begin{center}
\includegraphics[%
  width=.65\linewidth,
  keepaspectratio]{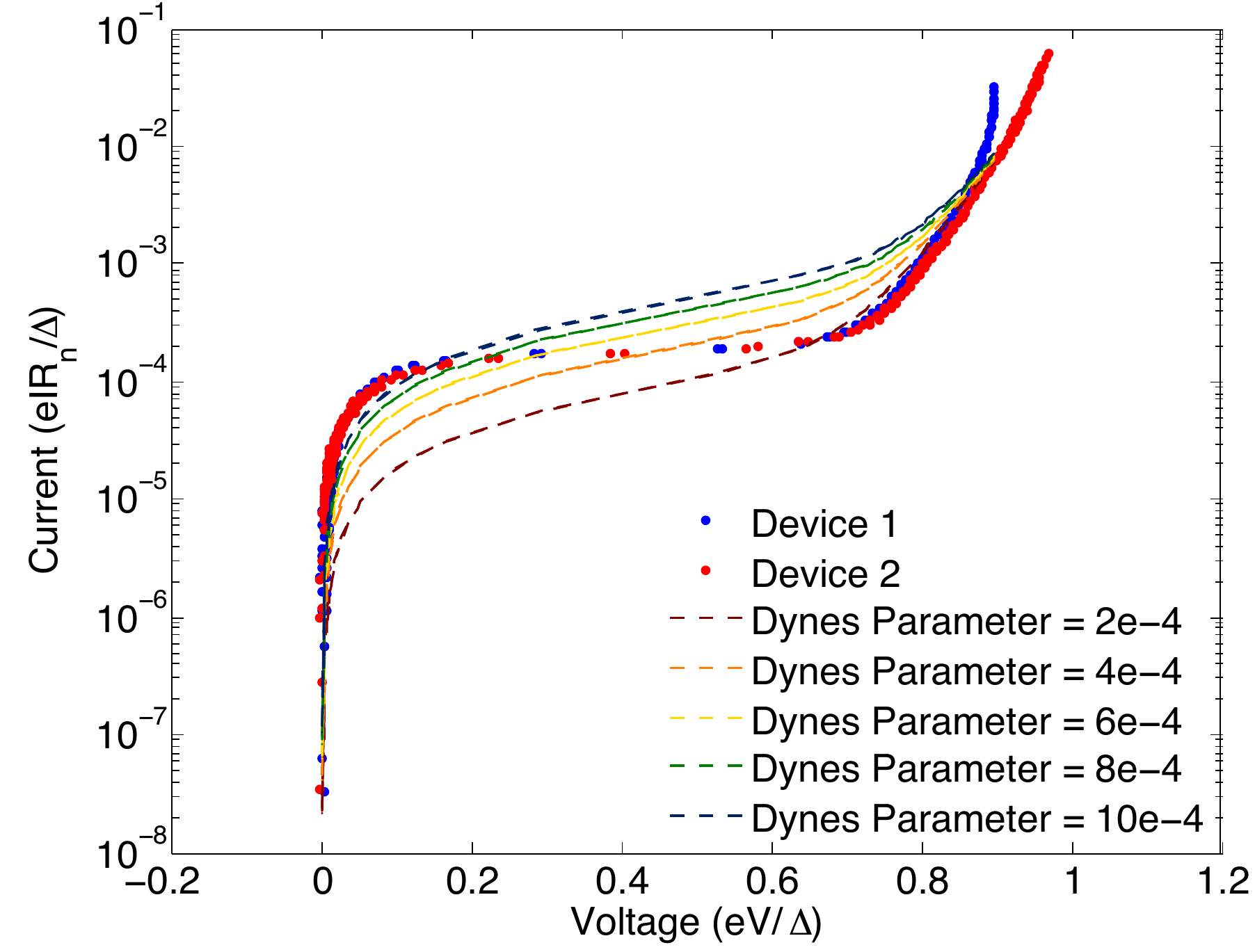}
\end{center}
\caption{(Color online) Current vs voltage for the first two devices plotted against NIS curves generated using various Dynes parameters.  The Dynes theory does not fit the data in the subgap.  The Dynes theory predicts a steadily rising current in the subgap, which we do not see in our data.  Instead, we see a flat plateau below the subgap, which is consistent with Andreev reflection.}
\label{fig:dynes}
\end{figure}

The results of our experiment are shown in Fig. \ref{fig:Andreev}.  The uncertainty in the measurement is $\pm$ 50 pA and $\pm$ 5 $\mu$V, which makes the error bars smaller than the markers in the plot.  The current was scaled by $eR_{n}/\Delta$, and the voltage was scaled by $e/\Delta$ in order to make the axes unitless.  The theory lines were calculated by using Eq. \ref{eq:andreev} with the measured device parameters shown in Table 1.  The divergence of the data with the theory at higher voltage biases, around 0.7 $\Delta/e$ and above, is due to the non-isothermal behavior of the NIS junctions.  In this work, we modeled the devices as being isothermal.  However, NIS junctions are known for their ability to cool electrons in the normal metal at bias voltages near the superconducting gap\cite{ClarkAPL}.  Therefore, in order to calculate the true behavior of these junctions, the IV curves need to be modeled by solving a complex power balance equation.  Since we are interested only in current due to Andreev reflections, which occurs below the gap, the junctions behave isothermally in the region of interest and no power balance modeling is required.  

As Fig. \ref{fig:Andreev} shows, our data are in excellent agreement with the theory below the superconducting gap.  Devices with the same $R_{SP}$ have the same dimensionless current, and the current due to Andreev reflections is inversely proportional to $R_{SP}$, as theory predicts.  Dividing the subgap data by the theory, Devices 1 and 2 agree with theory within 7 \%, Device 3 agrees within 18 \% and Device 4 agrees within 15 \%.  

For comparison, we also fit our data with NIS IV curves based on the phenomenological Dyne's parameter\cite{dynes}.  The Dynes parameter is added to the BCS density of states in an attempt to account for the broadening of the gap edge and/or the presence of subgap states, as shown in Eq. \ref{eq:dynes}.    

\begin{equation} 
\ I = \frac{1}{e R_{N}}\int_{0}^{\infty} \! [f_{N}(E-eV)-f_{N}(E+eV)] \left | \mathrm{Re} \left  [ \frac{E-i\Gamma}{\sqrt{(E-i\Gamma)^{2}-\Delta^{2}}}   \right ] \right | \mathrm{d} E
\label{eq:dynes}
\end{equation}

\noindent where $f_{N}$ is the Fermi function of the normal metal and $\Gamma$ is the Dynes parameter.

As Fig.  \ref{fig:dynes} shows, the Dynes theory does not provide a good fit to our data.  The Dynes theory predicts a steadily rising current below the gap, while in our data, the current plateaus, which is consistant with Hekking and Nazarov's theory.  This supports that we are measuring Andreev reflections and not subgap conductance due to the presence of subgap states.

\section{Conclusion}

In this paper, we have measured the current due to multiple Andreev reflections in NIS junctions and found that it is in excellent agreement with the theory presented by Hekking and Nazarov.  Their theory is valid only for voltage biases below the superconducting gap, and we are working to incorporate their theory with our power balance equations\cite{ClarkAPL} in order to make more accurate thermal models of NIS junctions.  These models will allow us to better predict the cooling properties of junctions with a small $R_{SP}$, for which the Andreev current is significant.


\begin{acknowledgements}
This work is supported by the NASA APRA program.
\end{acknowledgements}


\end{document}